# Magnetic-Moment Fragmentation and Monopole Crystallization

M. E. Brooks-Bartlett,[1] S. T. Banks,[1] L. D. C. Jaubert,[2] A. Harman-Clarke,[1,3] and P. C. W. Holdsworth[3,*]

[1]*Department of Chemistry, University College London, 20 Gordon Street, London WC1H 0AJ, United Kingdom*
[2]*OIST, Okinawa Institute of Science and Technology, Onna-son, Okinawa, 904-0495, Japan*
[3]*Laboratoire de Physique, École Normale Supérieure de Lyon, Université de Lyon, CNRS,
46 Allée d'Italie, 69364 Lyon Cedex 07, France*

The Coulomb phase, with its dipolar correlations and pinch-point–scattering patterns, is central to discussions of geometrically frustrated systems, from water ice to binary and mixed-valence alloys, as well as numerous examples of frustrated magnets. The emergent Coulomb phase of lattice-based systems has been associated with divergence-free fields and the absence of long-range order. Here, we go beyond this paradigm, demonstrating that a Coulomb phase can emerge naturally as a persistent fluctuating background in an otherwise ordered system. To explain this behavior, we introduce the concept of the fragmentation of the field of magnetic moments into two parts, one giving rise to a magnetic monopole crystal, the other a magnetic fluid with all the characteristics of an emergent Coulomb phase. Our theory is backed up by numerical simulations, and we discuss its importance with regard to the interpretation of a number of experimental results.

## I. INTRODUCTION

The interplay between frustration, topological order, fractionalization, and emergent physics has been the focus of a rapidly increasing body of work in recent years. In themselves, these concepts are not new. Frustration underpins theories of glassiness and has been much discussed since the seminal studies of Anderson and Villain [1,2]. Likewise, concepts of topological order and fractionalization, from the fractional quantum Hall effect [3] and quasiparticles in graphenelike systems [4] to solitons in one dimension [5,6], have been prominent topics in theoretical and experimental condensed-matter physics for over 30 years.

The subtleties of the interdependence between these concepts have been elucidated only recently, in the context of the so-called emergent Coulomb phase of highly frustrated magnetic models. However, the phenomenology of the Coulomb phase, with its characteristic dipolar correlations and emergent gauge structure, is not limited to the realms of frustrated magnetism. Indeed, similar behavior is observed in water ice [7] and models of heavy-fermion behavior in spinels [8], as well as in models of binary and mixed-valence alloys [9].

Henley [10] succinctly sets out three requirements for the emergence of a Coulomb phase on a lattice: (i) Each microscopic variable can be mapped onto a signed flux directed along a bond in a bipartite lattice, (ii) the sum of the incoming fluxes at each lattice vertex is zero, and (iii) the system has no long-range order. Elementary excitations out of the divergence-free manifold are seen to fractionalize, giving rise to effective magnetic monopoles that can interact via a Coulomb potential [11] and (in three dimensions) may be brought to infinite separation with finite energy cost.

In this paper, we demonstrate how the lattice-based magnetic Coulomb phase emerges in a considerably wider class of models than those covered by the conditions stipulated above. More specifically, we show that such a phase exists for a model "magnetolyte" irrespective of the magnetic monopole density or of monopole ordering [12]. We introduce the concept of magnetic-moment fragmentation, whereby the magnetic-moment field undergoes a novel form of fractionalization into two parts: a divergence-full part representing magnetic monopoles and a divergence-free part corresponding to the emergent Coulomb phase with independent and ergodic spin fluctuations [13]. Our results apply even for a monopole crystal that is shown to exist in juxtaposition with mobile spin degrees of freedom: a previously unseen coexistence between a spin liquid and long-range order induced by magnetic monopole crystallization. The implications of this field fragmentation are wide reaching and relevant for the interpretation of a number of experiments, as discussed below.

## II. THE MODEL

Magnetic monopoles [11] emerge as quasiparticle excitations from the ground-state configurations of the dumbbell model of spin ice. Here, the point dipoles of the dipolar spin-ice model [14] are extended to infinitesimally thin magnetic needles lying along the axes linking the centers of adjoining tetrahedra of the pyrochlore lattice (see Fig. 1) that constitute a diamond lattice of nodes for magnetic charge [15]. The needles touch at the diamond-lattice sites so that, by construction, the long-range parts of the dipolar interactions are perfectly screened for the ensemble of ice-rules states in which two needles point in and two out of each tetrahedron [16]. These ground states form a vacuum from which monopoles are excited by reversing the orientation of a needle, breaking the ice rules on a pair of neighboring sites. As vacuums go, this one is rather exceptional, as it is far from empty. Rather, the magnetic moments constitute the curl of a lattice gauge field [17], the Coulomb phase, with manifest experimental consequences for spin-ice materials. These effects include diffuse neutron-scattering patterns showing the sharp pinch-point features [18] characteristic of dipolar correlations, and the generation of large internal magnetic fields despite the status of vacuum [19,20].

---

[*]peter.holdsworth@ens-lyon.fr

In an analogy with electrostatics, the monopole-charge distribution obeys Gauss's law for the magnetic field $\vec{\nabla} \cdot \vec{H} = \rho_m$, where $\rho_m$ is the monopole density. The monopole number is not conserved, and the energetics of the dumbbell model at low temperature correspond to a Coulomb gas in the grand canonical ensemble, in which the phase space of monopole configurations is constrained by the underlying ice rules. One can define a Landau energy $\tilde{U} = U_C - \mu N$, where $U_C$ is the Coulomb energy of the neutral gas of $N$ monopoles and $\mu$ is the chemical potential such that $-2\mu$ is the energy cost of introducing a neutral pair of monopoles an infinite distance apart. For temperatures of order $-\mu$, the Coulomb-gas picture of spin ice must be extended to include doubly charged monopoles (see Appendix A).

The divergence in $\vec{H}$ is related to the breaking of the ice rules through $\vec{M}$, the magnetic-moment density that itself obeys Gauss's law $\vec{\nabla} \cdot \vec{M} = -\rho_m$. This constraint does not, however, define the entire moment density, which can have two contributions through a Helmholtz decomposition. The first $\vec{M}_m$ is the gradient of a scalar potential and provides the magnetic charge; the second $\vec{M}_d$ is a dipolar field, can be represented as the curl of a vector potential, and is divergence free [21,22]:

$$\vec{M} = \vec{M}_m + \vec{M}_d = \vec{\nabla}\psi(r) + \vec{\nabla} \wedge \vec{Q}. \quad (1)$$

In states obeying the ice rules, $\vec{M}_m = 0$, while $\vec{M}_d = \vec{M}$; this decomposition corresponds to an emergent Coulomb phase [10]. Breaking the ice rules and introducing magnetic charge lead to the conversion of $\vec{M}$ from the divergence-free $\vec{M}_d$ to the divergence-full field $\vec{M}_m$. Central to the discussions in this paper is the concept that this conversion is, in general, partial—the divergence-free part is not completely suppressed by the presence of the monopoles. Thus, in almost all circumstances, one has the coexistence of two complementary fields, reminiscent of the Hamiltonian splitting in topologically ordered phases [23].

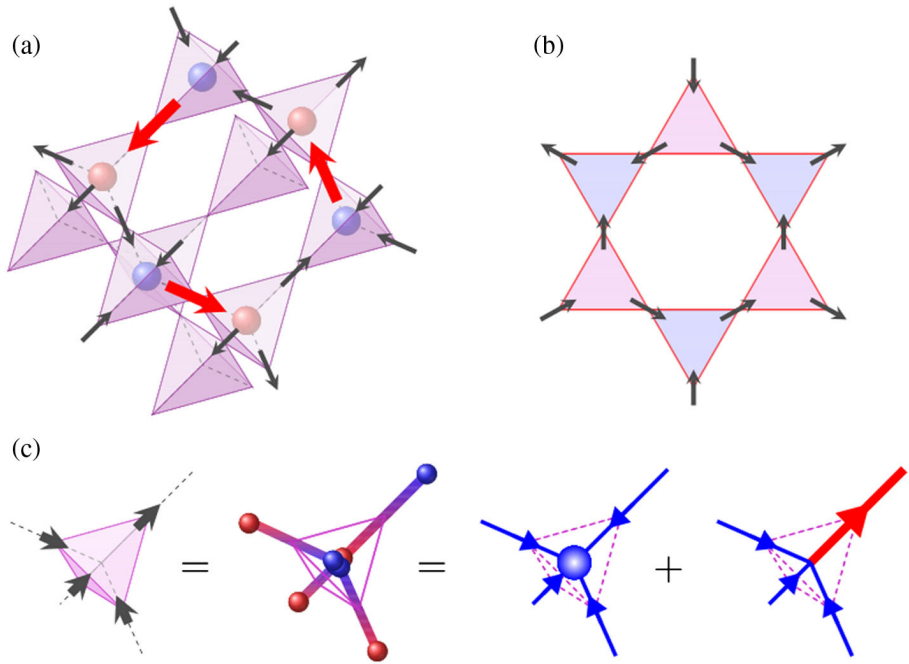

FIG. 1. Lattice structures: (a) Pyrochlore lattice, showing monopole-crystal and magnetic needle (spin) orientations. The sublattice index $\Delta_i$ defined in Eq. (4) is $+1$ (respectively, $-1$) on the diamond sites where the blue (respectively, red) monopoles sit. The minority spins (red arrows) are equivalent to dimers positioned on the diamond lattice. (b) Two-in–one-out spin configurations in the confined KII kagome spin-ice phase. (c) Spin and needle configurations for a three-in–one-out vertex carrying an isolated north pole showing magnetic-moment fractionalization into divergence-full and divergence-free elements. We emphasize that the divergence-free field emerging here, despite being a Coulomb phase, is different from the standard one in the absence of monopoles described in Ref. [11].

The total charge inside a microscopic volume surrounding a diamond-lattice site $i$ is $Q_i = -\int \vec{M} \cdot d\vec{S}$, with $d\vec{S}$ an outward-pointing surface element. For the dumbbell model, the integral reduces to a discrete sum $Q_i = -\sum_j M_{ij}$, where $M_{ij} = \vec{M} \cdot d\vec{S}_{ij} = \pm m/a$, $d\vec{S}_{ij}$ is an infinitesimal surface element cutting the needle, $m$ is the strength of the magnetic moment, and $a$ is the distance between diamond-lattice sites. The positive (negative) sign is for the needle pointing away from (into) the site $i$, so that $M_{ij} = -M_{ji}$. An ice-rules configuration with two needles into and two out of the site indeed gives $Q_i = 0$, while three in and one out (three out and one in) gives $Q_i = +(-)2m/a$ [11].

The fields for an isolated three-in–one-out vertex can be split into divergence-full and divergence-free parts subject to the constraint that the amplitude of each field element is $|M_{ij}| = m/a$:

$$[M_{ij}]\left(\frac{a}{m}\right) = (-1, -1, -1, 1)$$
$$= \left(-\frac{1}{2}, -\frac{1}{2}, -\frac{1}{2}, -\frac{1}{2}\right) + \left(-\frac{1}{2}, -\frac{1}{2}, -\frac{1}{2}, \frac{3}{2}\right). \quad (2)$$

The first set of fields satisfies Gauss's law for the charge at the origin; the second set satisfies a discrete divergence-free condition and constitutes a residual dipolar field dressing the monopole [see Fig. 1(c)]. Decomposition further away from the charge could be made by solving the lattice Poisson equation to find the field sets belonging to $\vec{M}_m$ that would be subtracted from the $M_{ij}$ to find the discrete elements of $\vec{M}_d$. Singly charged monopoles leave a residual contribution to $\vec{M}_d$ at each vertex, and it is only when the vertex is occupied by a doubly charged monopole for which $[M_{ij}] = \pm (m/a)(1, 1, 1, 1)$ that the contribution to $\vec{M}_d$ is totally suppressed. Hence, a fluid of singly charged monopoles should be accompanied by a correlated random dipolar field whose detailed structure is updated by the monopole dynamics and only destroyed on the temperature scale at which doubly charged monopoles proliferate. Indeed, the pinch points in diffuse neutron scattering from spin-ice materials are maintained up to surprisingly high temperatures [18,24], indicating the presence of such a dipolar field. The emergence of the dipolar field is further illustrated in Appendix C, where we show how the $[M_{ij}]$ are divided around a pair of isolated nearest-neighbor charges of the same sign.

The random fluctuations in the underlying gauge field can be ironed out by breaking the two-sublattice translational symmetry of the diamond lattice, creating a monopole crystal with north and south poles localized on different sublattices. For an ideally ordered array, the divergence-free fields on alternate sites are perfectly satisfied by the sets $[M_{ij}]_d = +(-)(m/a)(1/2, 1/2, 1/2, -3/2)$. Thus, one sees the emergence of a new Coulomb phase with extensive entropy superimposed on monopole order, in which each vertex has three contributions to the dipolar field of strength $1/2$ [in units of $(m/a)$] and one of strength $3/2$, which is shared between a pair of neighboring sites on opposite sublattices. This fragmented state could be termed an "antiferromagnetic Coulomb magnet" by analogy with the "ferromagnetic Coulomb magnet," predicted in the gauge mean-field theory of quantum magnets on a pyrochlore lattice [25]. The ordered component corresponds to a broken symmetry of the Ising spins described in the local-axis reference frame into the three-up–one-down or three-down–one-up sector. For the divergence-free part $\vec{M}_d$, placing a dimer along the bond of strength $3/2$ provides a mapping between the emergent dipolar field and hard-core dimers on the (bipartite) diamond lattice [26]. The extensive entropy of the dipolar field is thus associated with closed loops of dimer moves [27]. Introducing quantum-loop dynamics gives rise to a U(1) liquid phase close to the Rokhsar-Kivelson point [28,29].

A monopole-crystal ground state can be induced in the dumbbell model by modifying the chemical potential so that the total Coulomb energy $U_C$ outweighs the energy cost for creating the particles $-\mu N$. For the monopolar crystal $U_C = (N_0/2)\alpha u$, where $u = -\mu_0 Q^2/4\pi a$ is the Coulomb energy for a nearest-neighbor pair of monopoles of charge $\pm Q$, $\mu_0$ is the permeability, $N_0$ is the number of diamond-lattice sites, and $\alpha = 1.638$ is the Madelung constant. We define a reduced chemical potential $\mu^* = \mu/u$, and thus the ground state should be a monopole crystal for

$$\mu^* < \mu_0^* = \frac{\alpha}{2} = 0.819. \quad (3)$$

Monopole crystallization has also recently been studied within the dipolar spin-ice model, in the canonical ensemble, that is, with fixed monopole number [30], leading to a region of phase separation between the crystalline and the fluid phases. For classical spin ice, crossing this phase boundary would correspond to leaving the spin-ice phase [14], at which point double charges become favored. The ordering is then to a structure in complete analogy with zinc blende: a crystal of doubly charged monopoles for which $\vec{M}_d$ is everywhere zero, corresponding to the "all-in–all-out" (AIAO) magnetic order observed in $FeF_3$ [31].

In the modeling of spin ice, the chemical potential can be extracted from the dumbbell approximation to the dipolar spin-ice model (see the Supplementary Information of Ref. [11]). To put spin-ice materials in the context of the present work, we note that the dumbbell approximation for dysprosium titanate yields $\mu = -4.35$ K, while direct simulations for dipolar spin ice give $\mu = -4.46$ K [32], so that $\mu^* \approx 1.42$, well away from the monopole-crystal phase boundary. In Appendix A, we return to this modeling and show that the magnetostatics of monopoles leads to a prediction for the phase boundary where the ratio of nearest-neighbor exchange to dipolar energy scale reaches $J_{nn}/D_{nn} = -0.918$, in excellent agreement with direct analysis of the dipolar model [33]. Further, we show that, for a pair of doubly charged monopoles, both the chemical potential and the Coulomb energy are scaled by a factor of 4 compared with the values for the singly charged monopoles discussed in the main text. Hence, the zero-temperature spin-ice–monopole-crystal phase boundary occurs at the same point, whether one includes or excludes the doubly charged monopoles that complete the magnetic charge description of spin ice. In this paper, we explicitly exclude double charges, taking us away from traditional spin-ice modeling for large monopole concentrations, a point we return to in Sec. IV when discussing the experimental relevance of our results.

## III. RESULTS

We have tested these ideas directly through Monte Carlo simulations of the dumbbell model (details are given in Appendix B). In Fig. 2, we show the evolution of an order parameter for monopole crystallization $M_c$ and monopole number density $n$, as a function of reduced temperature $T^* = k_B T/|u|$ for different values of $\mu^*$. $M_c$ is defined as

$$M_c = \left\langle \left| \frac{1}{N_0} \sum_{i=1}^{N_0} q_i \Delta_i \right| \right\rangle, \quad (4)$$

where $q_i = (Q_i a / 2m) = \{-1, 0, +1\}$ is the topological charge on site $i$, $\Delta_i = \pm 1$ is a diamond-sublattice index (see Fig. 1), $\langle \ldots \rangle$ denotes a statistical average, and $n = \langle N \rangle / N_0$. The data show clear evidence of monopole crystallization at a transition temperature $T_C^*$ that varies with $\mu^*$. At this temperature, a lattice fluid gives way to a phase with reduced translational symmetry, in which $M_c$ approaches unity. Debye-Hückel theory for an unconstrained Coulomb gas on a bipartite lattice predicts a line of second-order transitions in the $(\mu^*, T)$ plane, becoming first order via a tricritical point as $\mu^*$ increases [34]. Our data are consistent with this scenario, despite the additional constraints of the dumbbell model. From the finite-size scaling analysis shown in Appendix E, we estimate a tricritical point $\mu_{tr}^* \approx 0.78$, $T_{tr}^* \approx 0.13$. This temperature is comparable to that obtained from the numerical simulation of a cubic-lattice Coulomb gas [35], although $\mu_{tr}^*$ is surprisingly close to $\mu_0^*$. For small values of $\mu^*$, a continuous transition takes the system from a high-density fluid ($n > 4/7$) to the crystalline phase. However, as $\mu^*$ increases toward the phase boundary at approximately 0.8, the fluid density is able to reach lower values near $T_c$ [shown by the data for $\mu^* = 0.794$ in Fig. 2(b)], indicating that while the crystalline ground state is energetically favored, the finite monopole density of the fluid phase is stabilized by entropy. It is this minimum in the density as the transition is approached that drives the transition to first order, as in the Blume-Capel model for spin-one systems [35]. More work is required to extract the effect of the constraints in detail and to establish the tricritical parameters with precision. There could, in principle, also be a liquid-gas transition at higher temperature, between low- and high-density fluids, but at the level of Debye-Hückel theory, this transition is suppressed by the monopole ordering [34]. There is no strong evidence of this liquid-gas transition in our simulations, although the crossover from a high-density fluid (region II of the phase diagram in Fig. 3) to a low-density fluid (region III) just outside the monopole-crystal phase boundary is quite sharp and could be considered as a vestige of such a transition.

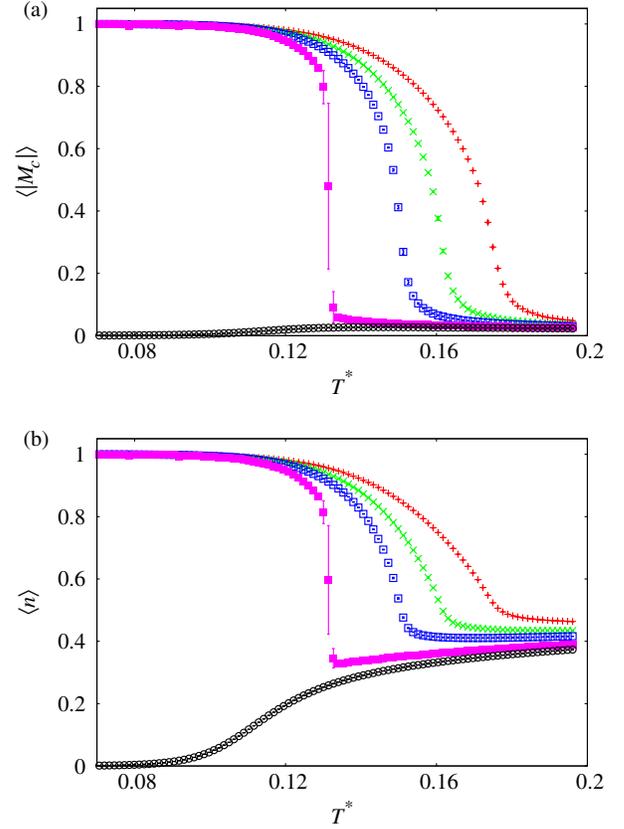

FIG. 2. (a) Order parameter $M_c$ for monopole crystallization, as defined in Eq. (4), and (b) density of charges $n$, as a function of reduced temperature $T^*$ for chemical potential $\mu^* = 0.767$ (+), 0.778 (×), 0.784 (□), 0.794 (■), and 0.801 (○). The transition appears to pass from second order to first order via a tricritical point for $\mu_{tr}^* \approx 0.78$. The very narrow first-order region ends at $\mu^* \approx 0.80$, above which spin-ice physics is recovered (e.g., the open black circles). The alternating positive or negative charges on the diamond lattice can also be seen as stacked monolayers of monopoles of the same charge in all three cubic directions. Note that the configurational constraints of the singly charged monopole fluid result in a high-temperature limit for the density of $n = 4/7$, rather than the $n = 0.5$ expected for an unconstrained bipartite lattice gas. A generalization of the worm algorithm has been developed to ensure equilibration at low temperature (Appendix B).

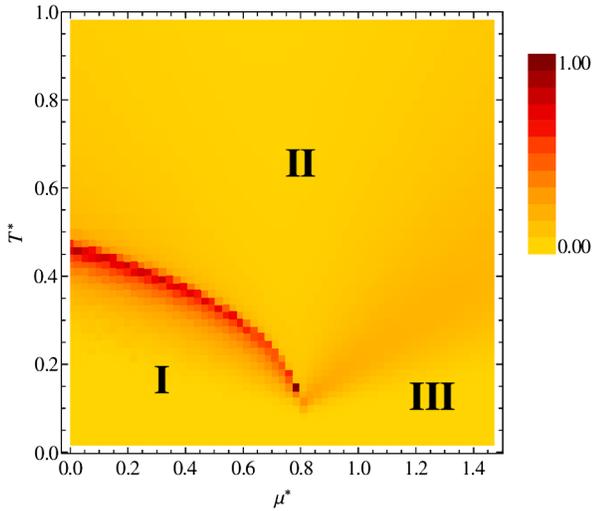

FIG. 3. The position of the singularity in the specific heat $C_\mu$ in the ($\mu^*$, $T^*$) plane traces out the phase boundary for the monopole-crystal phase. The most intense peak signals the passage of the transition from second to first order via a tricritical point. Indexes I, II, and III mark the monopole crystal, high-density fluid, and low-density fluids, respectively. Regions II and III are separated by a continuous crossover, and the gas of monopoles (III) corresponds to the low-temperature phase for spin ice.

In Fig. 3, we show the resulting phase diagram, as mapped out by the divergence of the specific heat $C_\mu$ at the phase transition. The monopole-crystal phase terminates for $\mu^* \approx 0.8$, in approximate agreement with our prediction of $\mu_0^* = 0.819$, the small difference being most likely due to finite-size effects exacerbated by the long-range interactions.

In Fig. 4, we show a simulated elastic neutron-scattering map determined within the static approximation, setting the magnetic form factor equal to unity and by averaging over 2000 randomly selected configurations of the ideal monopole crystal. For scattering purposes, the magnetic needles are once again taken as unit vectors (spins) $\vec{S}_i$ on the sites of the pyrochlore lattice. The structure factor $S(Q)$ is dominated by Bragg peaks at the (220) positions, characteristic of the all-in–all-out structure observed in FeF$_3$ [31]. The intensity of these peaks is precisely one quarter of that expected for complete all-in–all-out ordering and is consistent with scattering from monopoles constructed from fragmented spins with effective length 1/2. $S(Q)$ also reveals diffuse scattering with the clearly defined pinch points of a Coulomb phase. We have compared the intensity of the diffuse scattering with that found when the length of the minority spin at each vertex is extended to three while the length of the majority spins remains fixed at unity. The resulting structure factor has no Bragg peaks and has 4 times the intensity of the true diffuse scattering for a given wave vector. The ensemble of Bragg peaks plus emergent Coulomb phase therefore appears in excellent agreement with the predicted moment fragmentation of Eq. (2).

Magnetic charge crystallization also occurs for the dumbbell model on the kagome lattice (see Fig. 1) [36–40]. Breaking a $Z_2$ symmetry for needle orientations on each triangular vertex plunges the system from the unconstrained "KI" phase into the constrained "KII" Coulomb phase, realized in spin-ice materials by applying a field along the (111) direction of the pyrochlore lattice [41,42]. The mo-ments of a vertex with configurations constrained to two needles in and one out can, as above, be decomposed into divergence-full and divergence-free parts:

$$[M_{ij}]\left(\frac{a}{m}\right) = (-1, -1, 1)$$
$$= \left(-\frac{1}{3}, -\frac{1}{3}, -\frac{1}{3}\right) + \left(-\frac{2}{3}, -\frac{2}{3}, \frac{4}{3}\right). \quad (5)$$

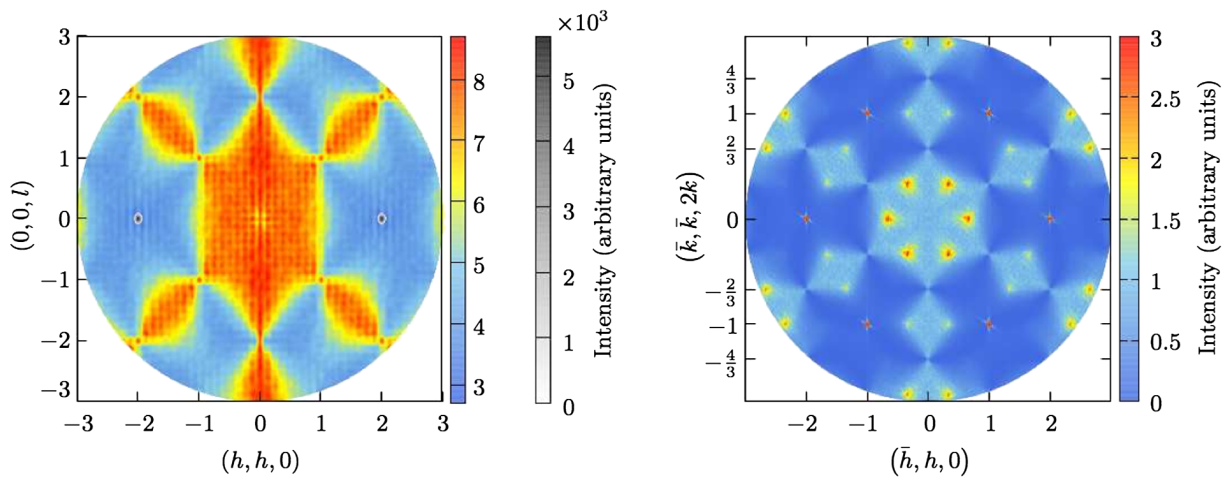

FIG. 4. Simulated unpolarized neutron-scattering structure factors $S(Q)$ for the pyrochlore monopole crystal (left) and for in-plane scattering from kagome ice (right). The pyrochlore $S(Q)$ has been calculated by averaging over 2000 distinct monopole-crystal ground states of a lattice with $L = 8$. In order to reveal the diffuse scattering, the Bragg peaks in the pyrochlore data are plotted as contours in gray scale superimposed on the contribution to $S(Q)$ from the dipolar field. The wave vector $Q$ is in units of $2\pi/a_0$, where $a_0 = 4a/\sqrt{3}$ is the lattice parameter of the cubic unit cell of the pyrochlore lattice. The kagome-ice data are taken from Ref. [102].

Consequently, simulated neutron-scattering plots for the ensemble of constrained states have both Bragg peaks and pinch points characteristic of a two-dimensional Coulomb phase (Fig. 4). The Bragg peaks have an intensity of $1/9$ of those for a fully ordered all-in–all-out phase. While deconfined monopole excitations away from these states carry a magnetic charge $Q = 2m/a$, as in spin ice, the charge ordering corresponds to a crystal of objects with charge $Q/2 = m/a$, providing a simple example of frustration-driven charge fractionalization [36–38,43].

Returning to the three-dimensional system, the persistent background fluctuations are further evidenced by studying local dynamics. We have collected two sets of data, one using local single spin-flip Metropolis dynamics, the other using a nonlocal worm algorithm [44] extended to include long-range interactions [45]. While the nonlocal algorithm is extremely powerful for extracting equilibrium properties in the highly constrained monopole-crystal phase, the local dynamics is of great interest, as it provides insight into how real systems might evolve with time [32,46]. The one-site, two-time monopole- and spin-autocorrelation functions are defined as

$$C_c(t) = \left\langle \frac{1}{nN_0} \sum_{i=1}^{N_0} q_i(0) q_i(t) \right\rangle, \quad (6)$$

$$C_s(t) = \left\langle \frac{1}{2N_0} \sum_{j=1}^{2N_0} \vec{S}_j(0) \cdot \vec{S}_j(t) \right\rangle. \quad (7)$$

In region I of our phase diagram, the charge-autocorrelation function $C_c(t)$ remains close to unity over the whole time window, reflecting the broken charge symmetry and the localization of monopoles, as shown in Fig. 5. The spin-autocorrelation function, however, shows a decay over a modest simulation time, from unity to an asymptote of $C_s(t = \infty) = 1/4$, reflecting the random projection of the spin onto the four orientations of a three-in–one-out or three-out–one-in vertex. This spin ergodicity, superimposed on a background of magnetic Bragg peaks, illustrates the collective nature of the monopole excitations: singular nodes in a fluctuating magnetic fluid, rather than static microscopic objects. We have evidence for the validity of these conclusions deep into region I, with simulations down to approximately $T_C/2$ for each $\mu^*$. $C_s(t)$ would be accessible experimentally, as it provides the diagonal contribution to the ac magnetic susceptibility

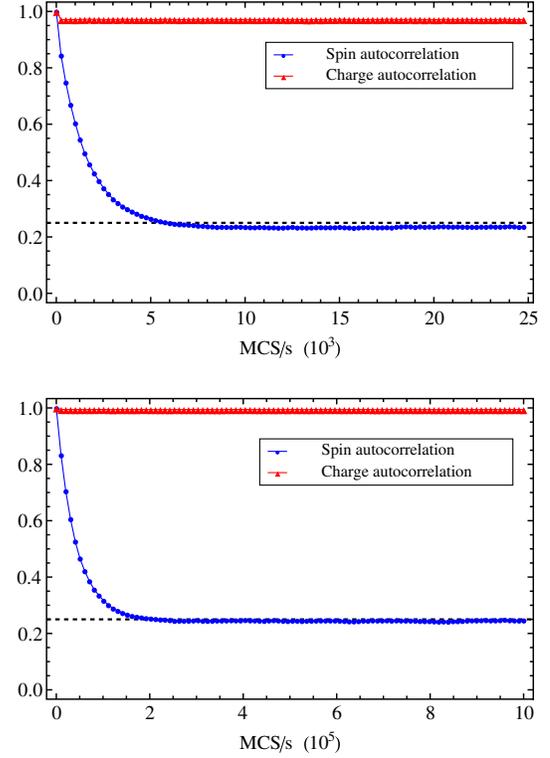

FIG. 5. Charge- and spin-autocorrelation functions for $\mu^* = 0.57$ (top) and $\mu^* = 0.41$ (bottom) for $T^* = 0.20$. The dotted line in each plot indicates the asymptote at $C_s(t = \infty) = 1/4$. The time required for the autocorrelation function to reach its longtime asymptote is significantly greater at $\mu^* = 0.41$ than at $\mu^* = 0.57$. This behavior is indicative of a slowing of the dynamics as the monopole density increases. As $\mu^*$ increases, $n$ decreases [see Fig. 2(b)] and the $t = \infty$ asymptote of $C_s(t)$ is slightly reduced relative to the value for the ideal monopole crystal. We note that both panels correspond to points far into region I.

$\chi(\omega)$, so that this finite time scale should show up as a characteristic frequency.

For the perfect monopole crystal, the dynamics are restricted to dimer-loop moves [27,47] if the system is to remain on the ground-state manifold. Single spin-flip dynamics lead to excitations away from these ground states, initially by monopole-pair annihilation. In region I, single spin-flip dynamics are dominated by needle flips that destroy and recreate nearest-neighbor pairs. Short-lived excitations of this kind collectively displace the fictive dimer positions approximating to the loop dynamics of dimers (Appendix D). The energy scale for an isolated excitation of this kind $d\tilde{U} \sim -u\alpha + 2\mu$ goes to zero at the ground-state phase boundary, allowing for extensive local dynamics in this region. Hence, for $M_c$ close to unity, the system is ergodic while at the same time retaining Coulomb-phase correlations.

We now turn to an important consequence of the field fragmentation for the interpretation of experimental data. The fluctuating background of the Coulomb phase appears to obscure the phase transition from bulk magnetic measurements. In Fig. 6(a), we show the magnetic susceptibility $\chi$ as a function of temperature for different values of $\mu^*$. At this level of analysis, the susceptibility is virtually featureless through the transition, showing no evidence of the characteristic cusp that one might expect at an antiferromagnetic transition in an Ising system. As one moves toward the tricritical point, a very weak feature does appear, driven by the huge monopole-density change as the system passes through the transition; however, the unusual characteristic (for a magnetic phase transition) of being virtually transparent to the bulk susceptibility remains essentially intact. A more detailed analysis of the susceptibility does, however, yield interesting information. It was recently shown that spin-ice models show a crossover in the Curie constant $C$ as the system moves from the uncorrelated high-temperature phase to the low-temperature Coulomb phase [15,48,49]. It was demonstrated that taking the needles as scatterers of unit length gives rise to a crossover from $C = 3T\chi = 1$ to $C \approx 2$. A similar Curie-law crossover is observed in the current work, as shown in Fig. 6(b), where $C$ is seen to evolve from $4/3$ at high temperature, as expected for a paramagnetic 14-vertex model, to a value $C \sim 3/2$ on entering the monopole-crystal phase and again to $C \approx 2$ on entering the constrained monopole vacuum. The change from a second- to a first-order transition can clearly be seen from this evolution. In the first-order region, $C$ evolves above $3/2$ as the monopole density drops in the fluid phase before falling discontinuously at the transition onto the $3/2$ plateau.

## IV. RELATION TO EXPERIMENT

One of our goals has been to construct a minimal model, based on spin ice, in which monopole order can be shown to coexist with (Coulomb-phase) spin-liquid physics. Spin ice is a central pillar in the ever-expanding field of frustrated magnetism. More generally, models of, and experiments on, systems based on pyrochlore and kagome lattices have resulted in a wealth of often puzzling results stemming from inherent geometrical frustration. We propose our model as a step toward answering some of these questions. The rest of this paper is dedicated to the consideration of experimental systems in relation to our model and its effects, including rare-earth oxides, spin-ice candidates, artificial spin ice, and the use of magnetic fields to induce a staggered chemical potential for magnetic charge.

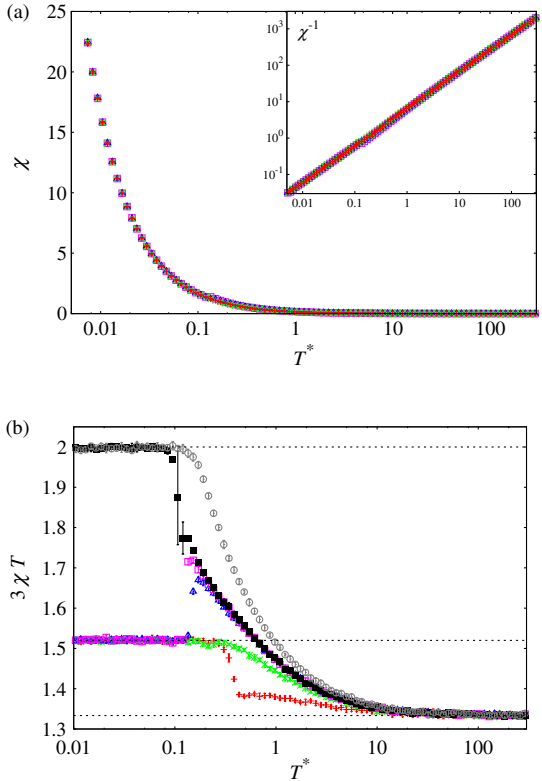

FIG. 6. (a) Magnetic susceptibility $\chi$ as a function of reduced temperature $T^*$ for chemical potential $\mu^* = 0.33$ (+), 0.65 (×), 0.784 (△), and 0.794 (□) $< \alpha/2$, where $\alpha$ is the Madelung constant. The inverse susceptibility is plotted in the inset on a log-log scale. (b) Curie-law crossover $3\chi T$ vs $T$ [same color labeling with two additional values of $\mu^* = 0.801$ (■) and 0.98 (○) $> \alpha/2$]. The dashed lines are theoretical expectations for the spin-liquid Curie-law prefactor $C$ of the Coulomb phase ($C \approx 2$) and the singly charged monopole fluid ($C = 4/3$), while the one for $C = 1.52$ is a guide to the eye. Equilibration has been ensured down to $T = 20$ mK ($T^* \approx 0.007$) by the worm algorithm. (See Appendix B for more details.)

### A. Magnetic field

An external magnetic field couples to both the monopoles and to $\vec{M}_d$, providing a gradient to the chemical potential, inducing transient monopole currents and ordering the dipolar field (the two processes being intimately related [11,15,46,50]). Applying a field in the [111] direction imposes kinetic constraints to monopole movement, restricting them to planes perpendicular to the field direction. In this scenario, the field constitutes a staggered chemical potential, breaking the $Z_2$ symmetry between the two sublattices, allowing experimental access to monopole crystallization [11]. This evolution corresponds to the first-order phase transition observed as spin-ice materials leave the plateau region of the $(\vec{B}, T)$ phase diagram. The corresponding phase boundary terminates at a critical end point $(\vec{B}_C, T_C)$ [51]. If one were to start from the charge-ordered Coulomb phase in zero field, an external (111) field would couple uniquely to $\vec{M}_d$. The system would then order via a three-dimensional Kasteleyn transition in complete analogy with that in two dimensions, driven by tilting the field off the (111) axis [52]. An external magnetic field can also stabilize a single-charge monopole in $Tb_2Ti_2O_7$ [53].

### B. Spin-ice candidates

Experimental relevance in zero field is a more open question. The chemical potential can be reduced within the confines of classical spin ice by reducing the lattice parameter, as is the case for the material $Dy_2Ge_2O_7$ [54]. Here, the monopole number certainly increases but their proliferation is accompanied by the generation of double monopoles with charge $\pm 2Q$. Close to the spin-ice, all-in–all-out phase boundary, the energy and entropy balance of the monopole crystal could possibly stabilize it ahead of either the monopole fluid or the fully ordered double-monopole crystal and it would certainly be interesting to study this problem both numerically and experimentally through high-pressure experiments or further substitution of smaller ions. Replacing germanium with silicon is, for example, a challenging possibility [55]. A further route to stabilization could be quantum fluctuations [25,56,57]. For systems close to the spin-ice–antiferromagnetic phase boundary [14], one might hope that zero-point fluctuations of the fragmented dipolar field could stabilize the monopole crystal over the classical all-in–all-out spin structure or the spin-ice manifold [28,29]. This situation would require both transverse spin fluctuations and dipole interactions between the moments, allowing the chemical potential to vary while permitting perturbative quantum spin fluctuations about the local $\langle 111 \rangle$ axes of spin ice. Another possibility is the generation of a staggered monopole chemical potential through a distortion of the lattice structure and the breaking of the crystal electric-field symmetry. Lifting the doublet degeneracy corresponding to the Ising-like spin-ice degrees of freedom in an ordered manner could thus lead to a perturbation that couples to monopoles but not to the dipolar field, favoring monopole crystallization.

### C. $Tb_2Ti_2O_7$

Fifteen years of intense research have made $Tb_2Ti_2O_7$ one of the most intriguing rare-earth frustrated magnets, sitting somewhere between a spin liquid [58] and quantum spin ice [59]—or maybe spanning both. It is this dual nature that makes it an interesting case study for magnetic-moment fragmentation. $Tb_2Ti_2O_7$ has a negative Curie-Weiss temperature $\Theta_{CW} = -14$ K [60]. As such, it can be considered as an antiferromagnet and numerical simulations of the corresponding dipolar spin-ice model give a phase transition to the all-in–all-out state at 1.2 K [14]. At ambient pressure, it fails to develop magnetic order down to at least 50 mK [61]. Recently, diffuse neutron scattering from single-crystal samples has exposed pinch-point–scattering patterns, indicating the presence of Coulomb-phase correlations [62,63], albeit somewhat deformed compared with those observed for classical spin-ice materials [18].

This picture is, however, very sensitive to perturbations. Under high pressure, the liquidlike phase transforms into a partially ordered structure where each vertex has one spin along a spin-ice axis and three collinear spins [64]. Vertices on the two-diamond sublattices have spin configurations with mirror symmetry, and the ordered state coexists with a fluctuating magnetic background. There is also evidence that a small magnetic field imposed along the [111] direction induces weak all-in–all-out order, leading to the material being described as an "incipient AIAO antiferromagnet" [65]. A strong field in the [110] direction stabilizes a double-layered structure of singly charged monopoles [53,66]. Recent experiments on polycrystalline samples have also shown that a very small amount of $Tb^{3+}$ stuffing produces an as yet unexplained long-range order, accompanied by weak antiferromagnetism [67]. Finally, both strong magnetoelastic coupling linked to quantum spin-liquid behavior [68,69] and splitting of the single-ion ground-state doublet [63,70–72] have been reported.

It would be naive to suggest a quantitative connection between the rich behavior of $Tb_2Ti_2O_7$ and the simple classical model presented here, but at a qualitative level, the similarities are striking. In the monopole-crystal phase, the model is an antiferromagnet with Coulomb-phase correlations. Placing it in the fluid phase but close to the monopole-crystal phase boundary, one could have residual pinch-point correlations from the underlying gauge field emanating from the correlated monopole fluid. Higher pressure could then send the system over the line into the crystalline state through a modification of the chemical potential [73,74]. The experimentally observed spin configuration in the high-pressure phase is quite different from the three-in–one-out spin-ice vertices of the monopole crystal, but it does share the same two-sublattice structure, to which one would have to add transverse spin relaxation off the spin-ice axes. The background of magnetic fluctuations is consistent with the fluctuating dipolar degrees of freedom $\vec{M}_d$. The incipient, or partial, all-in–all-out behavior in the presence of a field along [111] is exactly what one would expect of our model when sat close to the monopole-crystal phase boundary. As for the influence of the [110] field, since it does not give rise to monopoles at very low temperature in single crystals of $Ho_2Ti_2O_7$ or $Dy_2Ti_2O_7$ [75,76], the double-layer structure of monopoles observed in $Tb_2Ti_2O_7$ is a strong indication that such singly charged monopoles can be stabilized via internal couplings, even if the microscopic mechanism remains unknown. Spin-lattice coupling could be particularly relevant in this context [53,77].

Turning to dynamics, the freezing observed in $Tb_2Ti_2O_7$ [61,78,79] only involves a fraction of the spins ($\gtrsim 10\%$) and has been shown to be different from spin-glass physics [79]. Such partial spin freezing seems consistent with magnetic-moment fragmentation where only a fraction of the degrees of freedom order; the precise value of this fraction could then be mediated by quantum fluctuations. Hence, while microscopic modeling of $Tb_2Ti_2O_7$ is beyond the scope of this paper, we do propose magnetic-moment fragmentation as a promising route to understanding the apparent coexistence in this material of antiferromagnetism with the fluctuating Coulomb-phase physics of a frustrated ferromagnet [62].

### D. Other materials

A second quantum spin-ice candidate is $Yb_2Ti_2O_7$ [56,80]. This material, with a Curie-Weiss temperature estimated at around 600 mK [81], shows an unusual phase transition at 200 mK [82], with apparently no accompanying magnetic order [82,83] and magnetic dimensional reduction [84] in the high-temperature phase. There are, however, reports of a (partial) ferromagnetic ordering [84,85] at approximately 250 mK. The magnetic anisotropy in $Yb_2Ti_2O_7$ was initially considered to be $XY$-like [82], but more recent analysis has suggested that it could in fact be considered as a spin ice with the low-temperature behavior experimentally close to a quantum spin liquid [56,80]. Within this context, a quantum spin-liquid–classical spin-gas transition has recently been proposed [86]. With such complex behavior and sample dependence [87,88], a quantitative understanding requires a detailed microscopic approach [56,80,89]. That being said, the joint features of a low-temperature magnetically fluctuating phase and the presence of a phase transition that is partially transparent to magnetic probes are not unlike the fragmentation-driven transition in this paper, and the concepts developed here could be of use in understanding this complex material.

The understanding of $Tb_2Sn_2O_7$ also remains incomplete. This material orders in a ferromagnetic structure with spins canted off the local spin-ice axes [90,91], as predicted by combining dipolar interactions and spin relaxation [92]. However, the ordering is accompanied by an as yet unexplained fluctuating magnetic background, while the correlations above the transition appear to be antiferromagnetic. Although the details will almost certainly be different, magnetic-moment fragmentation does seem to be at play here and the concept could be of use in understanding this ordered, yet fluctuating, system.

### E. Artificial spin ice

There are immediate experimental consequences for our results for charge ordering in two dimensions. We have shown here that the KII phase on a kagome lattice, which was previously believed to be magnetically disordered, actually has partial all-in–all-out order [Eq. (5)]. Crystallites of the KII phase have recently been realized in permalloy nanoarrays with a honeycomb structure [93]. A simulated neutron-scattering analysis of the dipole orientations of the sample should therefore yield Bragg peaks of reduced intensity, similar to those observed in Fig. 4. Artificial spin-ice systems could therefore provide direct experimental realizations of magnetic-moment fragmentation. The (2, 2, 0) Bragg peaks characteristic of two-dimensional charge ordering should, in principle, also occur in spin-ice materials with a field along the (1, 1, 1) direction, although they may be masked by the field-induced magnetic order.

## V. CONCLUSIONS

In conclusion, we have shown how, through the presence of singly charged monopoles, a gauge field emerges from the dumbbell model of spin ice [11] that only partially maps onto the physical degrees of freedom, the magnetic needles. As a consequence, the intrinsic moments fragment into two parts. The first part satisfies the discrete Poisson equation on a diamond lattice, giving the magnetic monopoles, but does not exhaust the magnetic resources associated with each vertex. What remains forms an emergent dipolar field that evolves through monopole dynamics. By varying the chemical potential for monopole-pair creation, one can observe a monopole-crystallization transition, below which the gauge field provides a fluctuating and ergodic magnetic background with Coulomb-phase correlations. An analogous description exists for magnetic charge ordering in the KII phase of magnetic needles on a kagome lattice.

Order, or partial freezing in the presence of a fluctuating magnetic background, is a recurring phenomenon in frustrated magnetism (see, for example, Refs. [65,82,90,94,95]). Here, the magnetic-moment fragmentation leads naturally to persistent spin fluctuations within a purely classical model based on spin-ice physics. These background fluctuations mask the magnetic phase transition from view in susceptibility measurements, a phenomenon that can also occur in experiments on rare-earth pyrochlores [84,96,97]. It will be interesting to see if this concept of partial emergence can provide a more generic mechanism for persistent spin fluctuations in other situations. Finally, recent studies of quantum spin ices [25,98,99] have revealed a complete model for QED with magnetic monopoles, conjugate electric poles, and photon excitations. Including magnetic-moment decomposition within this model for QED opens the possibility for new levels of fractionalization, such as fractional charge and spin-charge separation.

## ACKNOWLEDGMENTS

We thank S. T. Bramwell for discussions concerning the simulated neutron-scattering plots and issues related to Ref. [22]. It is also a pleasure to thank O. Benton, M. Faulkner, T. Fennell, M. J. P. Gingras, J. Gardner, V. Kaiser, A. C. Maggs, P. McClarty, R. Moessner, and K. Penc for useful discussions. P. C. W. H. thanks the Institut Universitaire de France for financial support and is also grateful to the Yukawa Institute for Theoretical Physics, Kyoto, for hospitality during the workshop NQS2011.

S. T. B. thanks the École Normale Supérieure de Lyon for support and hospitality during a visit. M. E. B.-B. thanks the STINT Research Fund from Uppsala Universitet for financial support.

## APPENDIX A: THE COULOMB PHYSICS OF SPIN ICE

The dipolar spin-ice Hamiltonian can be written (see the Supplementary Information of Ref. [11]) within the dumbbell approximation as

$$H - H_0 = \frac{1}{2} \sum_{i \neq j} \frac{\mu_0 Q_i Q_j}{4\pi r_{ij}} + \frac{1}{2} v_0 \sum_i Q_i^2, \quad (A1)$$

where $Q_i$ is the total magnetic charge on diamond-lattice site $i$ and $v_0$ is an on-site term whose value is calculated from estimating spin-flip energies in the dipolar model. $H_0$ is the ground-state energy for a Pauling state within the dumbbell approximation $H_0 = -(N_0/2)v_0 Q^2$, with $Q = 2m/a$ the monopole charge. The ice rules and their consequent violation impose that $Q_i = 0, \pm Q, \pm 2Q$ only, and the diagonal term provides the chemical potentials for both singly ($\mu$) and doubly charged ($\mu_2$) monopoles:

$$\frac{1}{2} v_0 \sum_i Q_i^2 = -\mu N - \mu_2 N_2, \quad (A2)$$

where $\mu = -v_0 Q^2/2$, $\mu_2 = -2 v_0 Q^2$ and where the numbers of single and double monopoles are $N$ and $N_2$, respectively. The sketch below illustrates the energy scale, taking a single vertex or a diamond-lattice site from a two-in–two out ground state to a three-in–one-out monopole and finally to an all-in or all-out double monopole.

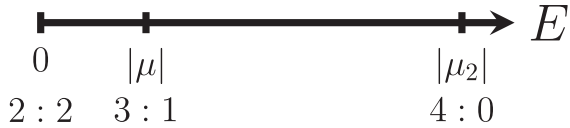

Neglecting the double monopoles, the internal energy $\tilde{U} = U_C - \mu N$ and we have the lattice Coulomb gas studied in the main text.

This analysis shows that both the chemical potential and the Coulomb energy scale by a factor of 4 when moving from single to double monopoles. Hence, the zero-temperature phase transition from monopole vacuum to monopole crystal [Eq. (3)] holds for both single- and double-monopole crystals, giving in both cases

$$v_0 = \frac{\mu_0 \alpha}{4\pi a}. \quad (A3)$$

Below this threshold, if both species are present, the excess Coulomb energy of the double monopoles wins, ensuring the predicted all-in–all-out ground state outside the classical spin-ice phase.

One can use the monopole crystallization as a criterion to estimate the position of the spin-ice phase boundary: Following Ref. [11],

$$|\mu| = \frac{v_0 Q^2}{2} = -\left[\frac{2J}{3} + \frac{8}{3}\left(1 + \sqrt{\frac{2}{3}}\right)D\right], \quad (A4)$$

where $J$ is the nearest-neighbor (antiferromagnetic) exchange constant and $D = \frac{\mu_0 m^2}{4\pi k_B a^3}$ is the strength of the dipole interaction between the moments of dipolar spin ice. The Coulomb interaction between nearest-neighbor monopoles can be written as $|u(a)| = \frac{8}{3}\sqrt{\frac{2}{3}}D$, so that Eq. (3) becomes

$$\frac{2J}{3} + \frac{8}{3}\left(1 + \sqrt{\frac{2}{3}}\right)D = \frac{\alpha}{2}\frac{8}{3}\sqrt{\frac{2}{3}}D, \quad (A5)$$

and hence

$$\frac{J_{nn}}{D_{nn}} = -\frac{4}{5}\left[1 + \sqrt{\frac{2}{3}}\left(1 - \frac{\alpha}{2}\right)\right] = -0.918, \quad (A6)$$

where $J_{nn} = J/3$ and $D_{nn} = 5D/3$. This Coulomb-gas estimate is in excellent agreement with numerical estimates for dipolar spin ice. Melko *et al.* [33] find $J_{nn}/D_{nn} = -0.905$ with hysteresis down to $J_{nn}/D_{nn} \simeq -1$, the origin of the difference being the small bandwidth for the Pauling states that is neglected in moving to the magnetic charge description.

In the present paper, the double charges are suppressed, leading to the monopole crystal with finite zero-point entropy and magnetic-moment fragmentation. In model spin ice, any perturbation that displaces the equality $\mu_2 = 4\mu$ in favor of single monopoles will generate the monopole-crystal phase in a band between the spin ice and the double-monopole-crystal phases. It is possible that the phase could be stabilized near this phase boundary by quantum fluctuations [100], thermal fluctuations, or a staggered, sublattice-dependent chemical potential. This point is addressed further in Sec. IV B, where we discuss the relevance of our work to experiment.

## APPENDIX B: SIMULATIONS

The numerical results in this paper are obtained by simulating the dumbbell model [11], equivalent to a gas of singly charged magnetic monopoles. The energy scale of the Coulomb interactions is entirely determined by $u(a)$, the energy scale for nearest-neighbor monopoles. The chemical potential $\mu^*$ is then a free parameter.

Three variants of Monte Carlo simulations are employed: (i) a single spin-flip Metropolis update (SSF), to reproduce the local dynamics relevant for classical spin-ice materials such as $Dy_2Ti_2O_7$ and $Ho_2Ti_2O_7$ (used for the results in Figs. 3 and 6); (ii) joint dynamics of SSF and a worm algorithm specifically designed for the current model (used for the results in Fig. 5; see below for more details on the worm algorithm); and (iii) SSF with worm updates and parallel tempering [101] (used for the results in Fig. 2).

The results presented in Fig. 3 are obtained using a system comprised of $8L^3 = 1000$ diamond-lattice sites where $L = 5$ is the number of cubic unit cells in each spatial dimension. The system is equilibrated over $t_{eq} = 10^4$ Monte Carlo steps per diamond-lattice site (MCS/s) with data collected over a further $10^5$ MCS/s.

For the results in Fig. 6, we use a system with $L = 7$ ($N_0 = 2744$) with $t_{eq} = 10^4$ MCS/s (for both values of $\mu^*$) after which observations of the autocorrelation function are made every MCS/s for a total of $2.5 \times 10^4$ MCS/s ($\mu^* = 0.57$) and every 100 MCS/s for a total of $10^6$ MCS/s ($\mu^* = 0.41$). Each point on these plots is the result of averaging over 100 consecutive observations. The density of monopoles $n$ in Eq. (5) is chosen as that at $t = 0$.

Figure 5 shows data obtained for a system with $L = 4$ ($N_0 = 512$). After annealing from high temperature to the temperature $T$ of interest over a period of $10^5$ MCS/s, the system is equilibrated at temperature $T$ for a further $t_{eq} = 10^5$ MCS/s prior to the data-collection period lasting $10^6$ MCS/s, during which observations are made every 10 MCS/s. Fifty worm updates are performed every 10 MCS/s to facilitate thermalization. We run six independent simulations for each value of the parameter $\mu^*$; the error bars are the standard deviations of these six samples at each temperature. Similarly, for the data in Fig. 2, the parameters are $L = 8$, $t_{eq} = 10^4$ MCS/s, with an observation period of $10^5$ MCS/s and averaging over four independent simulations. Again, 50 worm updates are performed every 10 MCS/s; 100 different temperatures between 0.2 and 0.6 K are used for parallel tempering.

### 1. Worm algorithm

In the absence of interactions between particles, the free energy of a system in the grand canonical ensemble only depends on the density of charges. With the addition of Coulomb interactions, the free energy also depends on the position of the charges. Hence, an update that

(i) does not modify the number of charges or their positions and
(ii) respects detailed balance, i.e., has the same probability flux to be formed and erased,

will necessarily be rejection free. In the absence of double charges, if we randomly choose an initial tetrahedron and spin (say, pointing "in"), it will always be possible to move forward and start a worm by flipping an "out" spin on the chosen tetrahedron. The number of out spins can be 1 (three in and one out), 2 (two in and two out), or 3 (three out and one in). Given that these choices remain the same irrespective of whether the worm is being created or destroyed, detailed balance is obeyed. When the worm closes on itself, it can be flipped at no energy cost while respecting detailed balance; the update can be accepted with probability 1. The strength of this algorithm is that it is rejection free in the Coulomb phase (two in and two out), in the dimer covering of the diamond lattice (alternating three in and one out and three out and one in), and for all densities of monopoles in between.

This worm algorithm could also be developed for the dipolar spin-ice model [14]. In this case, an additional global Metropolis argument would be needed to take into account the degeneracy lifting between states due to corrections of quadrupolar order when the needles of the dumbbell model are replaced by point dipoles on the nodes of the pyrochlore lattice.

### APPENDIX C: FIELD DISTRIBUTIONS

As an example of magnetic-moment fragmentation in the monopole fluid phase, we show, in Fig. 7, two isolated neighboring north poles on a square lattice. It is useful to consider this case (even though we do not consider in detail the dumbbell model on a square lattice [36] in this paper), as the fields can be easily visualized. Starting at nine o'clock and turning clockwise, the fields for sites 1 (on the left) and 2 (on the right) can be decomposed as follows:

$$[M_{ij}]^{(1)}\left(\frac{a}{m}\right) = (-1, -1, 1, -1)$$
$$= \left(-1, -\frac{1}{2}, 0, -\frac{1}{2}\right) + \left(0, -\frac{1}{2}, 1, -\frac{1}{2}\right), \quad (C1)$$

$$[M_{ij}]^{(2)}\left(\frac{a}{m}\right) = (-1, -1, -1, 1)$$
$$= \left(0, -\frac{1}{2}, -1, -\frac{1}{2}\right) + \left(-1, -\frac{1}{2}, 0, \frac{3}{2}\right), \quad (C2)$$

where in each case the first and second terms are the contributions to the divergence-full $\vec{M}_m$ and divergence-free $\vec{M}_d$ fields, respectively.

### APPENDIX D: THE DYNAMICS OF DIMER FLIPS

Single spin-flip Metropolis dynamics below the crystallization transition are dominated by needle flips that create and destroy north-south monopole charges. In Fig. 8, we show a typical sequence of moves for the monopole-crystal phase on a square lattice. In the first instance, the vertical needles flip independently, thus destroying the two neutral pairs of monopoles and reestablishing the ice rules on the four vertices of the square plaquette. These moves are followed by flipping of the horizontal needles that reestablishes locally the monopole crystal. A net consequence of such a sequence is to flip the fictive dimers from a horizontal to a vertical arrangement, as in a dimer-loop move [27]. Similar sequences occur for the pyrochlore and kagome lattices considered in this article, for which the shortest loop is a hexagon comprised of six needles or, equivalently, three dimers.

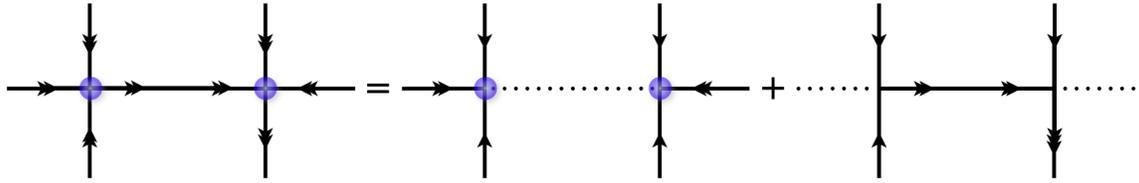

FIG. 7. Divergence-full and divergence-free field distributions for two isolated nearest-neighbor north poles (particle 1 on the left and 2 on the right) for the dumbbell model on a square lattice. Each chevron corresponds to a magnetic-moment field strength of $m/2a$ and a dotted line to zero-field strength. The blue circles represent north poles of charge $2m/a$.

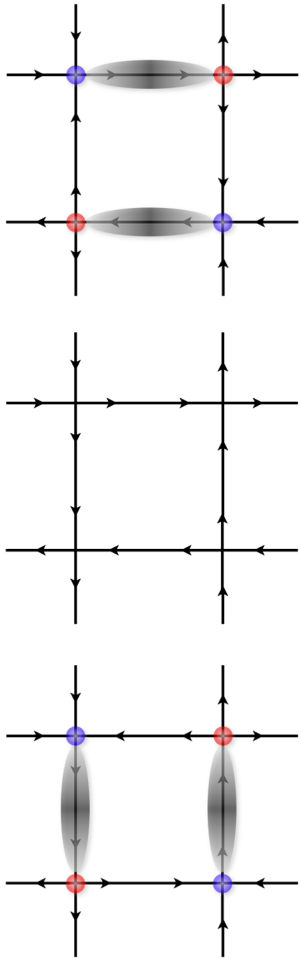

FIG. 8. A sequence of Metropolis updates in the monopole-crystal phase of a dumbbell model on a square lattice. The gray ellipses show the fictive dimer positions, and the blue and red circles signify north and south poles, respectively. The chevrons show the needle orientations along the bonds. The sequence of moves simulates a hard-core dimer flip on a square plaquette.

## APPENDIX E: FINITE-SIZE SCALING

The susceptibility of the monopole-crystallization order parameter is defined as

$$\chi_c = \frac{N_0(\langle m^2 \rangle - \langle m \rangle^2)}{k_B T}. \tag{E1}$$

In order to characterize more quantitatively the nature of the phase transitions, we perform finite-size scaling on the maxima of the specific heat $C_\mu$ and susceptibility $\chi_c$, plotted in Fig. 9. As explained in the paper, two regimes clearly appear. The transition is continuous up to a tricritical point at $\mu_{tr}^* = 0.78 \pm 0.01$, where it becomes first order before disappearing for $\mu^* > 0.800 \pm 0.05$. Our simulations suggest the continuous transition line to be of the 3D Ising universality class—consistent with a Debye screening of the long-range interactions—however, distinguishing between Ising and mean-field exponents is a difficult task [34] that would require further numerical and/or theoretical effort.

The discontinuity of the order parameter in Fig. 2 of the main text strongly supports the first-order nature of the phase transition, but its quantitative signature in finite-size scaling is rather challenging. That is, the diverging correlation length on either side of the tricritical point could lead to overestimates of $\mu_{tr}^*$ and $T_{tr}^*$. The first-order regime in our system occupies only a small region of parameter space, making it difficult to separate first-order from tricritical behavior. Increasing the system size beyond the correlation length rapidly becomes very time consuming—the long-range nature of the Coulomb interactions makes the CPU time scale as $L^6$. This computational cost is compounded by the relative inefficiency of the parallel tempering algorithm for first-order transitions, especially in large systems. Nonetheless, it has been possible to show a sharp increase of the scaling exponents close to the low-temperature phase boundary, their values approaching those of a first-order transition (see the lower panels of Fig. 9). In particular, the maximum in $C_\mu$ develops scaling behavior in this region—close to $\alpha/\nu = 3$ for $L \geq 4$ and $\mu^* = 0.796$.

An interesting consequence of our theory is the appearance of critical correlations even in the spin-ice regime where there is no phase transition. In Fig. 9, the green data points ($\mu^* = 0.801$) are constant for $L > 4$, as expected for a spin-ice crossover into the two-in–two-out Coulomb phase. However, for very small systems, both $C_\mu$ and $\chi_c$ seem to scale the same way as in the first-order region. This behavior suggests that spin-ice materials and models close enough to the low-temperature phase boundary can exhibit correlations inherited from the monopole crystallization, which should be visible using a local probe such as neutron scattering.

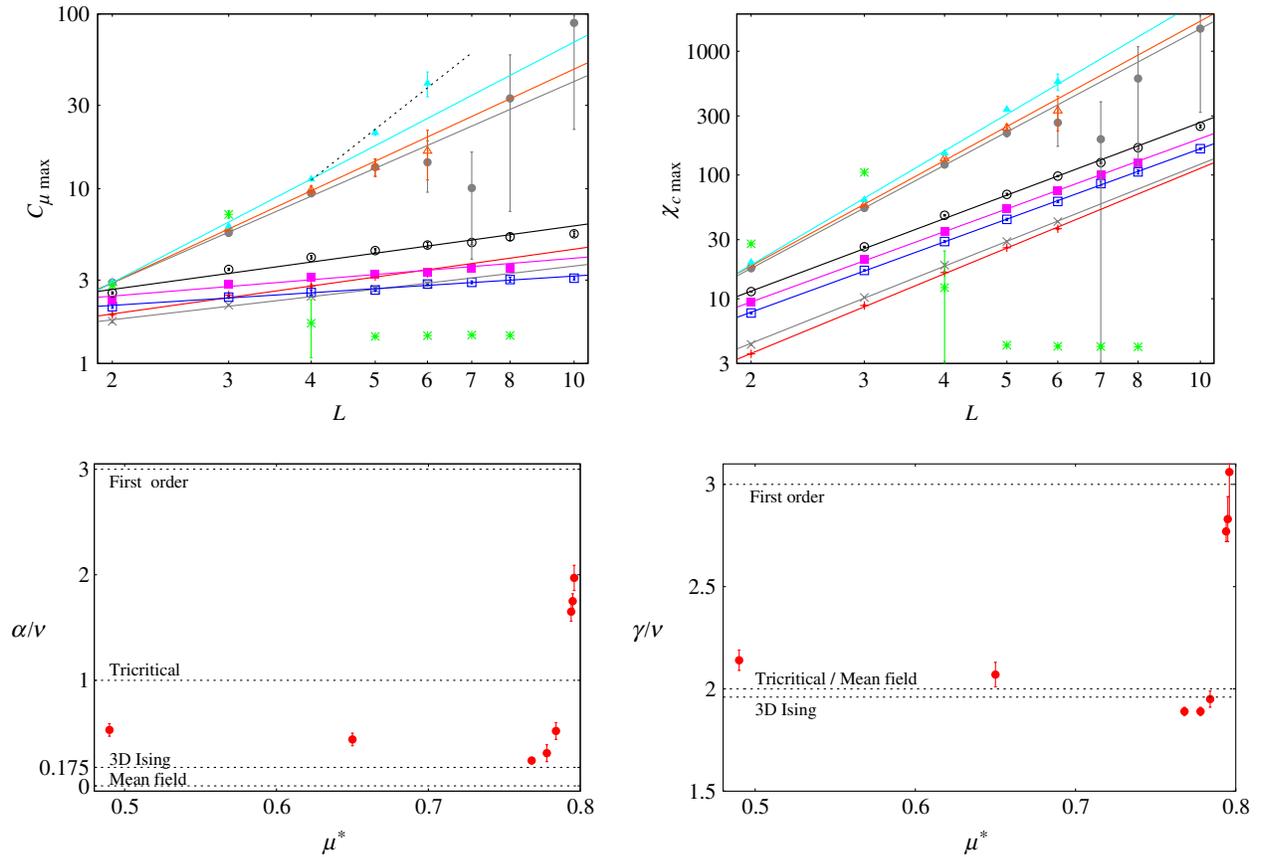

FIG. 9. Top: Finite-size scaling of the maxima of the specific heat $C_\mu$ and susceptibility $\chi_c$ of the order parameter $M_c$ as a function of linear system size $L$, for $\mu^* = 0.490$ (+), 0.654 (×), 0.768 (□), 0.778 (■), 0.784 (○), 0.794 (•), 0.795 (△), 0.796 (▲), and 0.801 (∗). The error bars are the standard deviation $\sigma$ over four independent simulation outcomes. Each solid line is the best fit obtained using Wolfram Mathematica v9.0 [103], including all data points for a given $\mu^*$ and weighting each data point by $1/\sigma^2$. The dashed line is a guide to the eye for the cubic power law ($\propto L^3$) appearing in the first-order regime for $L \geq 4$. Bottom: Scaling exponent ratios $\alpha/\nu$ and $\gamma/\nu$ as a function of $\mu^*$. The error bars represent a confidence level of 90%, based on the statistical uncertainty of the data plotted in the top panels.